\documentstyle[preprint,aps,prb]{revtex}
\tightenlines

\input epsf.sty
\begin{document}
\def\ang{\AA$\;$}


\title{Construction of Simulation Wavefunctions for Aqueous Species: 
D$_3$O$^+$}


\author{M. A. Gomez and L. R. Pratt}
\address{Theoretical Division
	Los Alamos National Laboratory, Los Alamos, New Mexico 87545
USA }

\author{LA-UR-98-2692}

\date{\today}

\maketitle
\begin{abstract}
This paper investigates Monte Carlo techniques for construction of
compact wavefunctions for the internal atomic motion of the D$_3$O$^+$
ion.  The polarization force field models of Stillinger, {\em et al.} and
of Ojamae, {\em et al.} were used. Initial pair product wavefunctions
were obtained from the asymptotic high temperature many-body density
matrix after contraction to atom pairs using Metropolis Monte Carlo.
Subsequent characterization shows these pair product wavefunctions to be
well optimized for atom pair correlations despite that fact that the
predicted zero point energies are too high.  The pair product
wavefunctions are suitable to use within variational Monte Carlo,
including excited states, and density matrix Monte Carlo calculations.
Together with the pair product wavefunctions, the traditional variational
theorem permits identification of wavefunction features with significant
potential for further optimization.  The most important explicit
correlation variable found for the D$_3$O$^+$ ion was the vector triple
product {\bf r}$_{OD1}\cdot$({\bf r}$_{OD2}\times${\bf r}$_{OD3}$).
Variational Monte Carlo with 9 of such explicitly correlated functions
yielded a ground state wavefunction with an error of 5-6\% in the zero
point energy.
\end{abstract}
\pagebreak

\section*{Introduction}

Flexible and dissociative models for simulation of liquid water have
often been used in classical statistical mechanical studies of aqueous
materials.
\cite{Stillinger:78,Morse:81,Christou:81,Stillinger:82,Reimers:82,%
Berens,Bopp:83,Jancso,Teleman:83,Thuraisingham:83,Deutsch,%
Wojcik,Wallqvist:90,Ruff,Slanina:90,Wallqvist:91,Zhu,Suhm:91,%
Ojamae:92b,Ojamae:92a,Corongiu:92,Smith:92,Halley,Corongiu:93a,%
Sciortino,Trokhymchuk:93,Vossen:94,Mizan,Duh,Kalinichev,David:96,%
Bansil,Silvestrelli,Rick,OSS} They present technical advantages for
carrying-out simulations and are of-the-essence where dissociation of
water molecules is necessary to the chemistry being studied.
Studies of clusters suggest that quantum mechanics plays a
non-negligible role in proton transfer.\cite{Tuckerman:97} However,
directly incorporating quantum mechanics via discretized path integral
approaches may require orders of magnitude larger computational  
effort than the classical problem. 
This Report investigates constructing
simple intramolecular wavefunctions for aqueous species.  These
wavefunctions might be used in Monte Carlo simulations of aqueous
solutions with the same computational techniques used to treat
flexible simulation models classically.  More specifically, our goal
is to determine the complication and accuracy to be expected in
constructing wavefunctions that might be transferrable and useful in
related studies of aqueous solution chemistry.

We take the deuterated hydronium ion, D$_3$O$^+$, as a specific
example species.  We choose this ion because of the substantial
current interest in proton exchange processes in water and in the
spectroscopy of charged clusters of water molecules; \cite{%
Stillinger:78,Halley,OSS,Tuckerman:97,Pomes:95,tuckerman:95,Pomes:96}
we choose the deuterated case to avoid the interesting complications
of spin restrictions on the wavefunctions.  In addition we will study
here only states of zero angular momentum, J=0.  Our goal will be to
obtain a simple intramolecular wavefunction for this species in a
reasonably organized fashion.

A traditional exhaustive expansion of the wavefunction into a basis
will produce satisfactory vibrational energies with effort but 
does not produce a simple result for the wavefunction.  Less
traditionally, diffusion Monte Carlo\cite{HLR} will produce
satisfactory vibrational energies for low energy states with effort
but not a simple wavefunction for other uses.

In contrast, the historical work of McMillan\cite{wlm}, a variational
Monte Carlo calculation for the ground state of liquid He$^4$, does
directly center on the construction of a simple wavefunction.  The
Monte Carlo character of this technique is limited to the primitive
but essential task of evaluating integrals with a many-body
wavefunction.  This approach permits simple but sophisticated
descriptions of correlation and can include a limited number of
excited states. The possibility of simultaneous treatment of excited
states provides an additional avenue for description of correlation.
In fact, the density matrix Monte Carlo approach
\cite{HLR,ceperley:88,bernu:90,Brown:95} of recent years can be
regarded as a generalization of the McMillan calculation; the
$\beta$=0 circumstance is precisely the result that we view as the
McMillan calculation with excited states.

The full power of the density matrix Monte Carlo approach can be used
with rough initial estimates of the wavefunctions sought.  The Monte
Carlo calculation systematically improves the computed energy levels.
However, since our goal is to derive a wavefunction useful in other
contexts, we emphasize finding and utilizing the best simple function
we can before the density matrix Monte Carlo procedure takes over.

We will initially use pair-product wavefuctions for atomic motion.  
The components of the pair-product are obtained by diagonalizing an 
approximate density matrix.  We emphasize that initial model  
vibrational wavefunctions need not be orthonormal.  

It is found (below) that these initial pair-product wavefunctions
capture two body correlations fairly well.  Using them as a basis in
McMillan and density matrix Monte Carlo calculations incorporates
further correlation effects.  To obtain more compact wavefunctions,
the variational principle is used to identify the important
many-body correlations.  These effects are included into the ground
state wavefunction and yield a significantly improved simple
wavefunction for the ground state.  A few of these correlated
wavefunctions can be used as the basis in a McMillan calculation
yielding reasonably accurate wavefunctions.

\section*{Model Wavefunctions for Atomic Motion}

We develop the vibrational wavefunctions analogously to development of
wavefunctions for liquid He. We initially seek a pair product form
\begin{eqnarray}
\label{pproduct}
\Phi_0 = \prod_{OD~pairs} \phi_{OD}(r_{OD}) \prod_{DD~pairs}
\phi_{DD}(r_{DD})\enspace.
\end{eqnarray}
In this case, we expect the functions $\phi_{OD}$ to be local mode
orbitals and the functions $\phi_{DD}$ to serve as correlation
functions.

The casual interpretation of these functions as orbitals will serve to
distinguish vibrational excitations.  The method sketched here is
based upon the observation that the eigenfunctions of the asymptotic
$\beta \rightarrow 0$ density matrix are suitable initial estimates of
the pair functions sought.  This method has the additional advantage 
that harmonic analysis of a potentially rough energy landscape is not 
required.

We start with a function of this simple form for several reasons.
First, a similar distribution is implied when flexible and
dissociative pair simulation models of water are treated classically.
Second, elements of such a form might be approximately transferrable
to OD and DD joint distributions in other settings - this would be the
natural assumption for simulation calculations.  Third, such a form
would be similar, though not the same as a ground state vibrational
wavefunction for a harmonic system.  The distinctions are that normal
modes coordinates are not used, and that functions employed can attempt
to treat anharmonic systems by deviating from Gaussian form where the
potential is anharmonic.  The remainder of this paper characterizes
such wavefunctions and investigates some improvements.

\subsection*{Natural orbitals for general oscillators in one dimension}
The natural orbitals of a density 
matrix $ \langle x'; \beta \vert x \rangle \equiv
<x'|e^{-\beta H}|x>$ are introduced by  
\begin{equation}
\int \langle x'; \beta \vert x \rangle \phi_{n}(x')dx' = 
e^{-\beta E_{n}} 
\phi_{n}(x) \enspace.
\label{norbs}
\end{equation}
In fact the asymptotic $\beta \rightarrow 0$ limit of the density
matrix is simple:\cite{feynman}
\begin{equation}
 \langle x'; \beta \vert x \rangle = (m/2 \pi \beta\hbar^2)^{1/2}
\exp\{-m(x-x')^2/2\beta\hbar^2-\beta(V(x)+V(x'))/2\}~.
\end{equation}
Sethia {\it et al.\/} showed that though the density matrix used is
approximate, satisfactory orbital functions are obtained.\cite{Singh}
This form for the thermal density matrix is more than merely an
approximation.  For small $\beta$, it is asymptotically correct.
Therefore, it is satisfactory if the application permits a small
$\beta$.  The subsequent developments exploit this point.

\subsection*{Contraction to Pairs for  the many-body Thermal Density 
Matrix}

For higher dimensional problems, we can exploit the approximate
density matrix to produce the necessary reasonable pair functions
after tracing-out all other degrees of freedom.  This tracing-out will
be generally possible on the basis of classical Monte Carlo
techniques.

Consider a system composed of N particles located at $({\bf
r}_{1},{\bf r}_{2}\ldots{\bf r}_{N})$.  The N-body high-temperature
thermal density matrix is
\begin{eqnarray}
\label{r0}
\langle{\bf r}_1,{\bf r}_2 \ldots {\bf r}_N;\beta\vert{\bf r'}_1,{\bf
r'}_2\ldots{\bf r'}_N\rangle & = & \prod_{k=1}^N (m_k / 2 \pi \beta
\hbar^2)^{3/ 2}
\exp\{-m_k({\bf r}_k-{\bf r'}_k)^2/{2\beta\hbar^2} \}  \nonumber \\ 
 & \times & \exp\{-\beta(V({\bf r}_1,{\bf r}_2\ldots{\bf r}_N) + V({\bf
r'}_1,{\bf r'}_2\ldots {\bf r'}_N))/2\}\enspace.
\end{eqnarray}
We now focus on the \{ij\} pair of particles. For all other particles,
we restrict the density matrix to the diagonal.  For example, taking
\{ij\}=\{12\} we obtain
\begin{eqnarray}
\label{rdiag}
\langle{\bf r}_1,{\bf r}_2 \ldots {\bf r}_N;\beta\vert{\bf r'}_1,{\bf
r'}_2\ldots{\bf r}_N\rangle & \propto & \prod_{k=1,2}(m_k/ 2 \pi \beta
\hbar^2)^{3/ 2}
\exp\{-m_k({\bf r}_k-{\bf r'}_k)^2/{2\beta\hbar^2} \} \nonumber \\ 
 & \times &  \exp\{-\beta(V({\bf r}_1,{\bf
r}_2\ldots{\bf r}_N) + V({\bf r'}_1,{\bf r'}_2\ldots {\bf r}_N))/2\}.
\end{eqnarray}
Now for the \{12\} pair we transform to the \{12\} center of mass and
relative coordinates: ${\bf R} = (m_1 {\bf r}_1 + m_2 {\bf r}_2)/(m_1 +
m_2) $ and ${\bf r} = {\bf r}_1 - {\bf r}_2 $.
\begin{eqnarray}
\label{rdiagtrans}
\lefteqn{ \langle {\bf R}+m_2{\bf r}/M,{\bf R}-m_1{\bf r}/M \ldots
{\bf r}_N;\beta \vert {\bf R'}+m_2{\bf r'}/M,{\bf R'}-m_1{\bf
r'}/M\ldots{\bf r}_N\rangle \propto } \nonumber \\ \hfill & \hfill &
\exp\{-\beta(V({\bf R}+m_2{\bf r}/M,{\bf R}-m_1{\bf r}/M\ldots{\bf
r}_N) + V({\bf R'}+m_2{\bf r'}/M,{\bf R'}-m_1{\bf r'}/M\ldots {\bf
r}_N))/2\} \nonumber
\\ \hfill &  \times & (M/2 \pi \beta \hbar^2)^{3/ 2} \exp\{-M({\bf
R}-{\bf R'})^2/{2\beta\hbar^2} \} (\mu/ 2 \pi \beta \hbar^2)^{3/ 2}
\exp\{-\mu({\bf r}-{\bf r'})^2/ 2\beta\hbar^2 \},
\end{eqnarray}
where $M=m_1+m_2$ and $\mu = m_1 m_2/M$.  

Note that $R$ is not the molecular center of mass and that the
potential generally does depend on $R$.  However, in the interest of
simplicity and in view of the form sought Eq. (\ref{pproduct}), we
will trace-out the $R$ dependence also.  Our final step will be to
use classical Monte Carlo techniques for the integrations required to
bring Eq. (\ref{rdiagtrans}) into the form of Eq.(\ref{norbs}) for the
relative coordinate r.  Thus we sample configurations with the
probability density $ \exp\{-\beta V({\bf r}_{1},{\bf r}_{2}\ldots{\bf
r}_{N}) \} $ and we estimate the kernel
\begin{eqnarray}
\label{kernel}
K({\bf r},{\bf r'}) & \propto & \left< e^{-\mu({\bf r}_1-{\bf
r}_2-{\bf r'})^2/2\beta\hbar^2 } \delta({\bf r}-{\bf r}_1+{\bf
r}_2)e^{-\beta
\delta V({\bf R}+m_2{\bf r'}/M,{\bf R}-m_1{\bf r'}/M \ldots {\bf
r}_N)/2} \right>_{e^{-\beta V}},
\end{eqnarray}
where
\begin{eqnarray}
\label{ddV}
\delta V({\bf R}+m_2{\bf r}'/M,{\bf R}-m_1{\bf r}'/M\ldots{\bf r}_N) &
\equiv & \nonumber \\ V(\left[ m_1{\bf r}_1+m_2{\bf r}_2 +m_2{\bf
r'}\right]/M,\left[ m_1{\bf r}_1+m_2{\bf r}_2 -m_1{\bf
r'}\right]/M\ldots{\bf r}_N) & - & V({\bf r}_1,{\bf r}_2\ldots{\bf
r}_N)\enspace. 
\end{eqnarray}
$K({\bf r},{\bf r'})$ is symmetric.
For the case considered
here, it does not depend on the angles that {\bf r} or {\bf r}$'$ make
with the laboratory fixed coordinate system. Thus, this kernel depends
only on r, r$'$, and an angle.  Since we here focus on
radial functions, we only need the kernel after having averaged
over that polar angle.

Note that for determination of the orbitals only, the normalization of
the kernel $K({\bf r},{\bf r}')$ is not significant. To within an
unimportant normalization constant we can evaluate the required kernel
through the following procedure:
\begin{enumerate}
\item Draw configurations $({\bf r}_1, {\bf r}_2 \ldots {\bf r}_N) $
from the probability distribution proportional to $ \exp\{-\beta
V({\bf r}_{1},{\bf r}_{2}\ldots{\bf r}_{N}) \} $ utilizing the
Metropolis Monte Carlo algorithm\cite{metropolis,kalos}.
\item For each configuration, choose an $r'$ from a grid and its 
corresponding spherical angles $\theta'$ and $\phi'$ using
quasi-random number series.  Weight each configuration $({\bf r}_1,
{\bf r}_2
\ldots {\bf r}_N) $ and the corresponding ${\bf r'}$ by
\begin{eqnarray}
\label{sample}
e^{-\mu({\bf r}_1 - {\bf r}_2 -{\bf r'})^2/ 2\beta\hbar^2 } e^{- \beta
\delta V( \left[ m_1{\bf r}_1+m_2{\bf r}_2 +m_2{\bf r'}\right]/M,\left[
m_1{\bf r}_1+m_2{\bf r}_2 -m_1{\bf r'}\right]/M \ldots {\bf r}_N))/2}
\end{eqnarray}
\item Perform a final integration over angles
\begin{eqnarray}
{\overline K}(r,r') \propto \int_{0}^{\pi} d\theta \int_{0}^{\pi}
d\theta'
\int_{0}^{2\pi} d\phi' 
\int_{0}^{2\pi} d\phi ~\sin \theta ~K({\bf r},{\bf r'}) ~\sin\theta'
\end{eqnarray}
$\theta , \phi , \theta' $, and $\phi'$ are the spherical, angles
corresponding to ${\bf r}$ and ${\bf r'}$.  For this case of $J=0$, it
would have been sufficient to integrate over the angle between ${\bf
r}$ and ${\bf r'}$.
\end{enumerate}
With this kernel in hand, we solve the one dimensional equation
\begin{eqnarray}
\label{eigen}
\int_0^\infty {\overline K}(R,R')R'^2 \phi(R')dr' = \lambda \phi(R)~.
\end{eqnarray} $\lambda$ is not $e^{-\beta E}$ but is proportional to 
it.

This approach is not limited to the circumstances that one member of
that pair is a massive molecular center.  This can be applied also for
DD pairs in the D$_3$O$^+$ molecule.  However, for the D$_3$O$^+$
molecule in particular, it is natural to regard the OD functions as
local mode orbitals and the DD functions as providing a subsequent
account of correlations.

\section*{D$_{3}$O$^{+}$ Pair Functions}

The functions $\phi_{OD}$ and $\phi_{DD}$ are found for the
second version of the Ojamae-Shavitt-Singer (OSS) potential\cite{OSS} using
$\beta^{-1}$=0.01 Hartree and a sample size of 500,000 $({\bf r}_1,
{\bf r}_2 \ldots {\bf r}_N)$ configurations.  This energy parameter is
6.3 kcal/mole (more than 3000 K), a value above the inversion
barrier of 4.4 kcal/mole. Smaller values for $\beta$ are problematical
because of the fragmentation of the D$_3$O$^{+}$ molecule at high
temperatures.

\begin{figure} 
\hspace{0.7in}\epsfbox{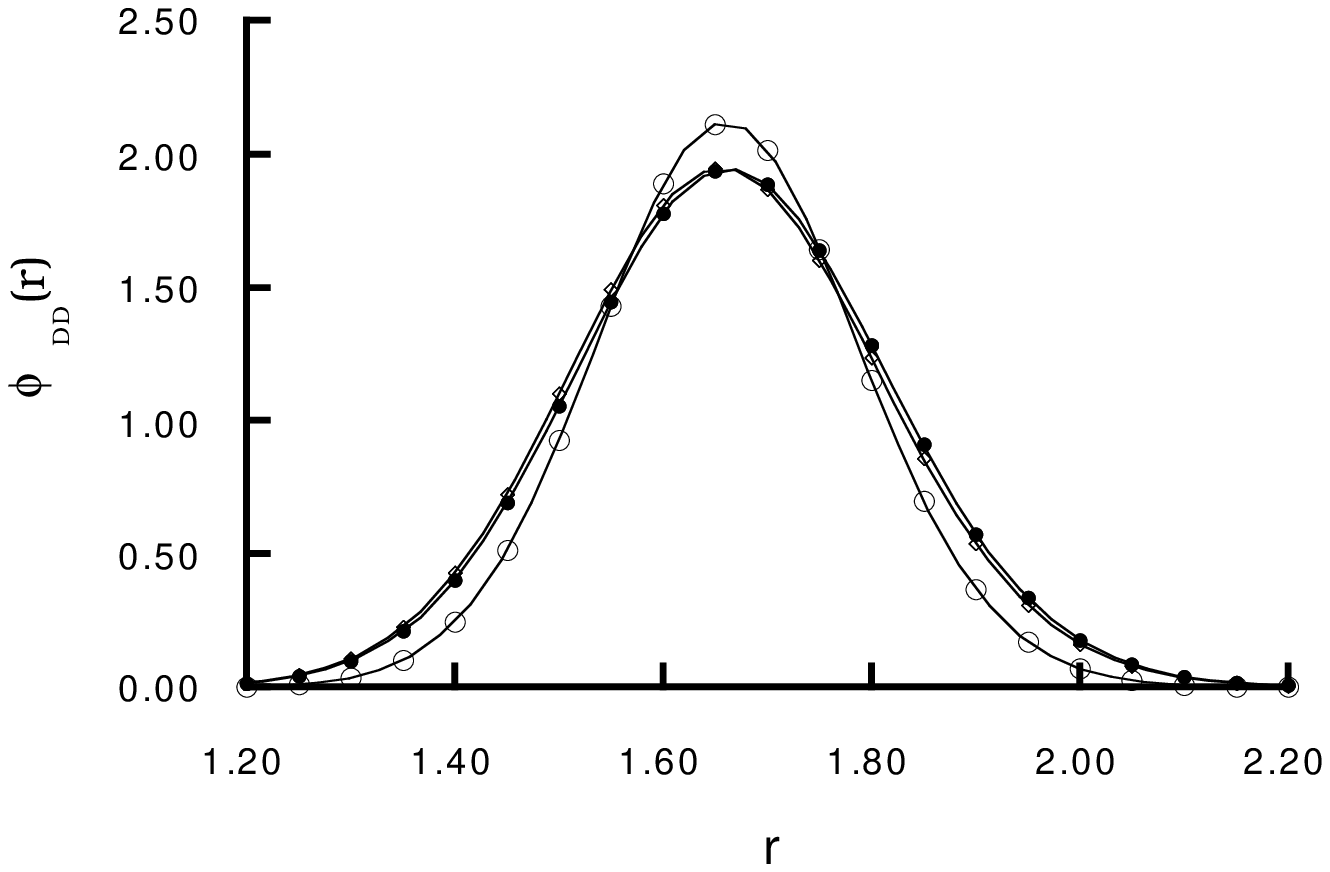}
\vspace{0.1in}
\caption{The $\phi_{\rm DD}(r)$ obtained by fixing one of the 
deuterons (line with open circles) is significantly narrower 
than that obtained from unconstrained sampling (line with diamonds).
The co-linear assumption of fixing the orientation of the reference 
bond (line with solid circles) is not significant.  The points were 
fit to cubic splines.  Displacements are in \ang.}
\label{phiDD} 
\end{figure}

Figure~\ref{phiDD} shows $\phi_{DD}$ obtained.  The additional
approximation of fixing one of the deuterons in the reference bond has
the significant effect of narrowing the pair function.  As expected,
the effect is not as dramatic for fixing an oxygen atom.  The
co-linear assumption of fixing the orientation of the reference bond
is not significant.  

\begin{figure} 
\hspace{0.7in}\epsfbox{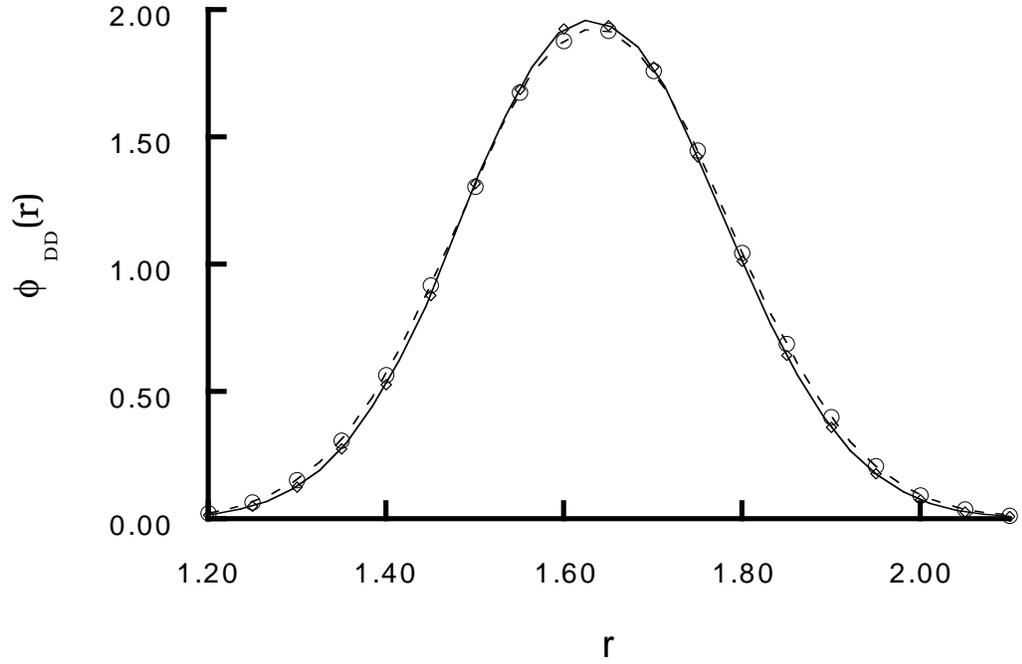}
\vspace{0.1in}
\caption{The $\phi_{\rm DD}(r)$ obtained using
$1/\beta$=0.01677$\dots$~Hartree (open circles on dashed line) is slightly 
wider than that obtained using $1/\beta$=0.01~Hartree (diamonds on 
solid line).  The points were fit to cubic splines.
Displacements are in \ang.}
\label{TauComp}
\end{figure}

These functions were also obtained for the Stillinger, Stillinger, 
Hogdgon (SSH) potential.\cite{SSHpot} A comparison of features of SSH 
and OSS is presented in Table~\ref{CompPot}.  The higher (more 
realistic) binding energy for the additional proton in the SSH case 
and the higher inversion barrier of this potential permits a smaller 
$\beta$.  In that case, the pair functions are found for two values of 
$1/\beta$, namely 0.01~Hartree and 0.01677$\ldots$~Hartree.  The 
difference in the ``frequencies'' associated with the ground and first 
excited state OD ``local mode'' functions changed from 3150~cm$^{-1}$ to 
1987~cm$^{-1}$.  The corresponding differences in DD frequencies 
changed from 1243~cm$^{-1}$ to 837~cm$^{-1}$.  In spite of these large 
frequency changes, the pair functions, seen in Figure~\ref{TauComp}, 
are qualitatively similar.

\section*{Accuracy of the Pair Product Wavefunctions}

The numerical arrays describing the OD and DD functions were fit to 
cubic splines with first derivatives set to zero at the end 
points\cite{NumericalRecipies} and the pair product wavefunctions of 
Eq.~(\ref{pproduct}) were obtained.  Several steps were taken to 
assess the quality of these wavefunctions.  For the OSS and SSH 
potentials, the variational Monte Carlo\cite{wlm} zero point energies 
of the pair product wavefunctions are 0.0321(2) and 
0.0367(6)~Hartrees, respectively.  Simple diffusion Monte Carlo\cite{HLR} 
zero-point energies for the potentials are 0.02516(3) and 
0.03042(3)~Hartrees.  (All errors are two standard deviations unless 
otherwise noted.)  The energy for the pair product function is 
between 20\% and 30\% too large.  This function was further optimized by 
introducing parameters to move (m) and scale (s) the pair functions; 
$\phi(r_{new})=\phi((r_{old}+m)/s)$.  No significant lowering of the 
energy was found.

The quality of optimization of these pair functions can be analyzed 
directly.  The usual variation principle can be expressed as
\begin{equation}
\left \langle  \left( \delta \ln \Psi \right) E_L(1..N) \right\rangle_{\left| \Psi \right|^2} 
- \left\langle 	\left( \delta \ln \Psi \right)
\right\rangle_{\left| \Psi \right|^2} 	\left\langle E_L(1..N)
\right\rangle_{\left| \Psi \right|^2}=0~,
\label{variationalprinciple}
\end{equation}
where $E_L=\Psi^{-1}H\Psi$ is the usual local energy function and the
subscripted brackets $\langle \ldots \rangle_{\left| \Psi \right|^2}$
indicate an average with probability density $\left| \Psi
\right|^2$.   For a pair product wavefunction this form implies
\begin{equation}
1 -{{{\left\langle {\hat \rho _{\gamma \nu }^{(2)}(r)E_L(1..N)}
\right\rangle_{\left| \Psi \right|^2}}\over {\left\langle {\hat \rho
_{\gamma \nu }^{(2)}(r)} \right\rangle_{\left| \Psi \right|^2}}
{\left\langle {E_L(1..N)} \right\rangle_{\left| \Psi \right|^2}}}}
=0~,
\label{variationalCondition}
\end{equation}
where $\hat \rho _{\gamma \nu }^{(2)}(r)$ is the $\gamma\nu$ pair
joint density operator.

This stationary requirement is satisfied for the exact ground state
wavefunction since the local energy is spatially constant.  
Deviation of the left side of Eqn.~\ref{variationalCondition} from
zero will be defined as optimization deficit.

\begin{figure}
\hspace{0.4in}\epsfbox{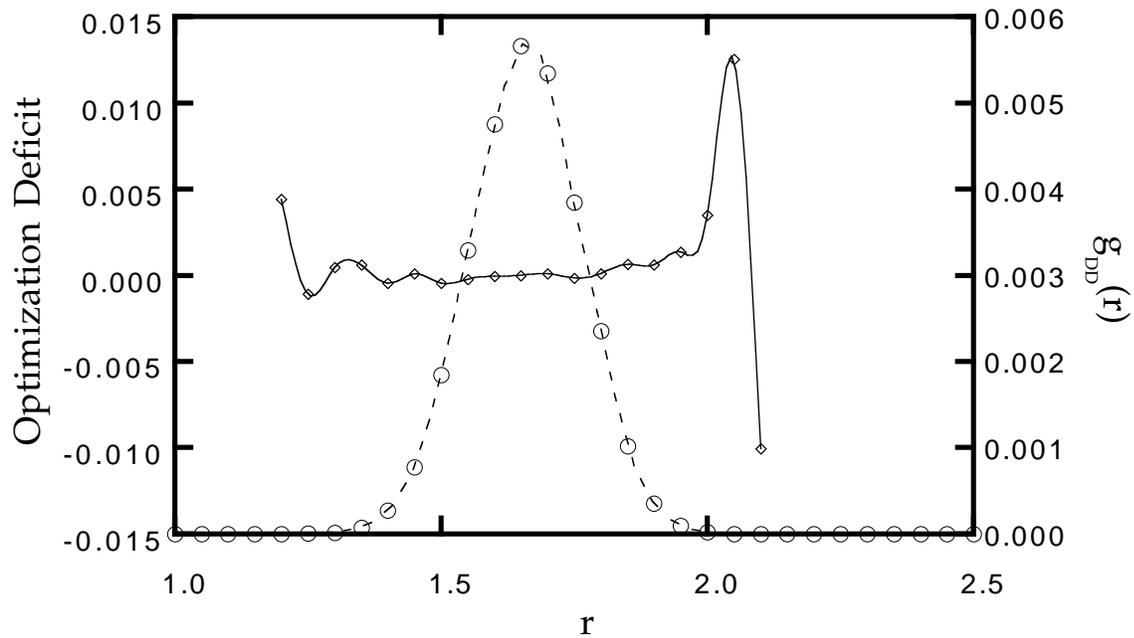}
\caption{DD optimization deficit (diamonds on solid line) and 
VMC radial distribution function (open circles on dashed line)
for OSS model.}
\label{ErrorDD}
\end{figure}

Figure~\ref{ErrorDD} shows the optimization deficit for the OSS DD 
pair functions and the DD VMC radial distribution function.  The OD 
functions show similar low optimization deficits; these pair 
functions lead to accurately uncorrelated pair densities and local 
energies and are thus well optimized.  Note that the largest deficits 
are at the tails of the distribution.  Thus further optimization would 
require significant effort because of the difficulty of getting 
statistically significant data for those infrequent events.

\section*{Correlation beyond Pairs:  Variational and Density Matrix 
Monte Carlo}

We can use model excited state functions to improve the ground state
and get a few low lying states at the same time.  A concise expression
of these approaches is obtained by focusing on the density matrix.
Suppose we have a number of approximate wavefunctions $\vert \Phi_i
\rangle, i = 0, \ldots, N$.  By Monte Carlo procedures detailed
elsewhere \cite{ceperley:88,bernu:90}, we estimate the matrix elements
\begin{eqnarray}
\label{cbapproach:a}
\overline {\rho}_{ij} (\beta) \equiv <\Phi_i|\exp\{-\beta H\}|\Phi_j>,
\end{eqnarray}
and 
\begin{eqnarray}
\label{cbapproach:b}
\overline {H}_{ij}(\beta) \equiv <\Phi_i|H\exp\{-\beta H\}|\Phi_j>,
\end{eqnarray}
for $\beta$ not too large.  As $\beta$ gets large, information is lost
for highly excited states, so this approach depends on location of an
intermediate range of $\beta$.  $\overline H$ and $\overline \rho$ are
then diagonalized simultaneously.  The $\vert \Phi_i \rangle$ need not
be orthonormal initially since Monte Carlo methods will be used to
perform the required integrations.  The $\beta\rightarrow$ 0 limit
may be viewed as a direct extension of the McMillan\cite{wlm} approach.
The trace of ${\overline \rho}_{ij}$ provides a natural importance
function for the Monte Carlo calculation.

To apply these methods to our problem, an ordered set of excited state 
functions is constructed.  The ratio of eigenvalues of the 
$n^{\rm th}$ and $0^{\rm th}$ eigenfunctions in Eq.~\ref{eigen} are 
used to assign relative energies to the pair product functions.  
Model excited state functions are built-up by linearly combining
functions representing higher levels of excitation.

These functions were used in density matrix Monte Carlo calculations.  
In its $\beta=0$ limit, density matrix Monte Carlo mixes the lowest 
energy functions much like configuration interaction in electronic 
structure calculations.  Tables~\ref{OSSCFMCtau=0} 
and~\ref{SSHCFMCtau=0} show the energies obtained in these variational 
calculations for OSS and SSH potentials, respectively.  The guiding 
function used was $ \Psi_{\rm trial}(q) = \left(\sum_{i=0}^{M} 
\phi_{i}(q)^{2}\right)^{\nu}$ with M=5 and $\nu={{1}\over{4}}$.  
$\nu$ of ${{1}\over{4}}$ is used instead of ${{1}\over{2}}$ so that 
higher states than M are efficiently sampled.  Alternatively, M could 
be made larger.  However, the cost of making M larger is higher than 
the cost of making $\nu ={{1}\over{4}}$.  Independent estimates for 
the zero-point energy of $0.02516(3)$~Hartree ($5522(7)$~/cm) and 
$0.03042(3)$~Hartree ($6676(7)$~/cm) are found using diffusion Monte 
Carlo for the OSS and SSH potentials, respectively.  Inclusion of a 
small number of excited states substantially improves the predicted 
zero-point energy; the 20-30\% error with the pair product 
wavefunction is reduced to about a 10\% error with inclusion of nine 
states.  However, the improvement of this prediction with increasing 
numbers of excited states is slow.  As shown in 
Tables~\ref{OSSCFMCtau=0} and~\ref{SSHCFMCtau=0}, inclusion of 94 and 
99 states still results in a 4-6\% error.

\begin{figure}
\hspace{0.4in}\epsfbox{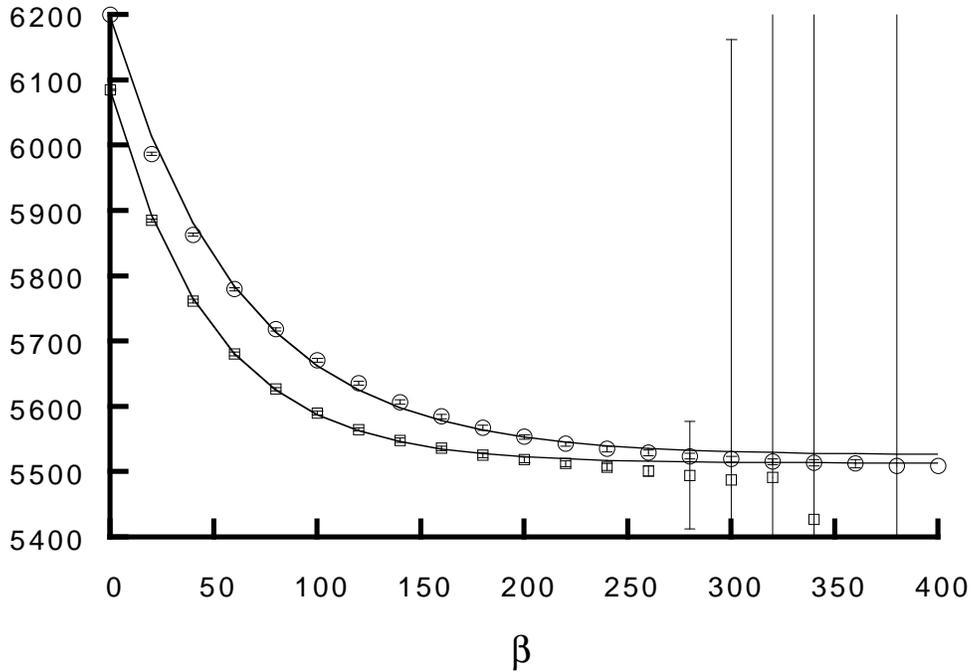}
\caption{Convergence of the lowest energy state to the ground state 
energy as a function of $\beta$ for the OSS potential.  The circles 
and squares were obtained using 9 and 17 states, 
respectively.  The lines are Marquardt's fits.}
\label{OSSground9}
\end{figure}

The variational energies are significantly improved by density matrix
Monte Carlo as $\beta$ increases from zero with 9 excited state
functions.  Including more functions accelerates the convergence.
(see Figure~\ref{OSSground9}) Tables~\ref{OSSCFMCtauplateau}
and~\ref{SSHCFMCtauplateau} show the energy estimates for lowest 5
symmetric states for OSS and SSH respectively.  The average energies
and their standard deviations as a function of $\beta$ were used in a
Marquardt's fit to the function $A+Be^{-{\beta}/C}$.  Error bars were
obtained by dividing the configurations used to calculate the matrices
into smaller sets, diagonalizing these sets, obtaining the standard
deviation for their eigenvalues, and dividing it by the square root of
the number of sets.  Assuming that the distribution of errors is
Gaussian, the ellipsoid representing 90\% error can be
plotted.\cite{NumericalRecipies} The errors reported in
Tables~\ref{OSSCFMCtauplateau} and~\ref{SSHCFMCtauplateau} are
$\chi^{2}$ 90\% confidence intervals.  Some data was not well
approximated by a decaying exponential.  In this case, the point in
the plateau with the smallest standard deviation is given.  For this
data which is designated by a *, 2 standard deviations are given in
parenthesis.

Because it is relevant to assessment of potential energy surface
models, we note in passing that for the OSS potential the excited
vibrational state with a node {\bf r}$_{OD1}\cdot$({\bf
r}$_{OD2}\times${\bf r}$_{OD3})$=0 has an energy of 0.02522(4)~Hartree
as determined by a fixed node diffusion Monte Carlo calculation.  This
is about about $15(24)$ /cm higher than the zero point energy.  A
model fit to experiment by Sears {\em et al.} yielded a gap of 15/cm.
However, the OSS potential overestimates the inversion barrier and,
therefore, we expect it to underestimate this splitting.  Beyond this
comparison, our results for both the OSS and SSH potential differ
signicantly from the results of Sears {\em et al.} For example, first
totally symmetric excited states are 794(10)/cm and 731(11)/cm above
the ground state for OSS and SSH respectively.  The first totally
symmetric excited state in Sears' fit was 453/cm above the ground
state.\cite{Sears}

\section*{Expressing Correlations Beyond Pairs More Compactly}

The variational principle Eq.~\ref{variationalprinciple} and the
optimization deficit can be used to search for structural variables
that are correlated with the local energy.  Such variables are
principal candidates for construction of explicitly correlated
wavefunctions.

\begin{figure} 
\hspace{0.7in}\epsfbox{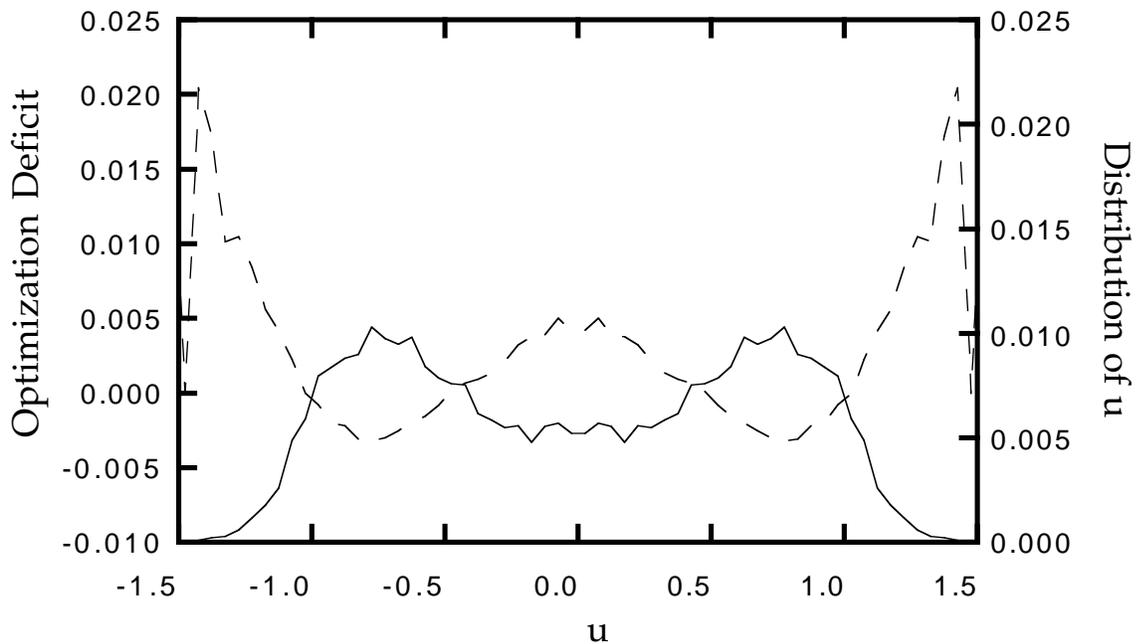}
\vspace{0.1in}
\caption{The OSS VMC distribution of u (solid line) and the 
optimization deficit (dashed line) 
is shown for the pair product wavefunction}
\label{3body} 
\end{figure}

\begin{figure} 
\hspace{0.7in}\epsfbox{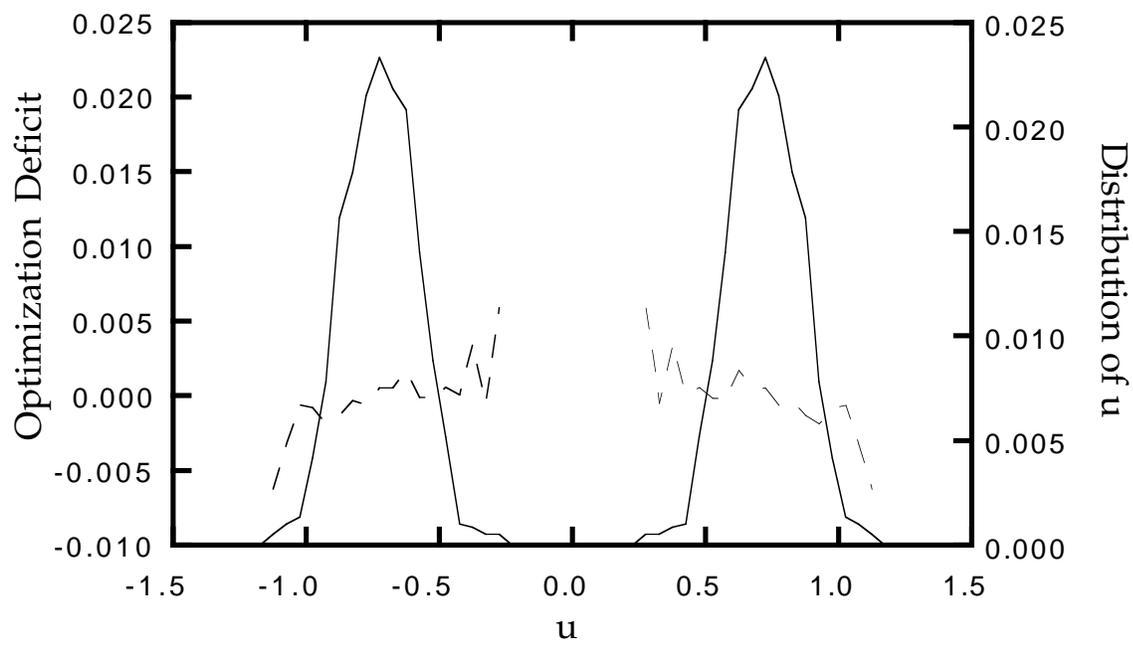}
\vspace{0.1in}
\caption{The OSS VMC distribution of u (solid line) and the 
optimization deficit (dashed line) is shown
for the correlated optimized wavefunction}
\label{3bodyOpt}
\end{figure}

Consideration of several natural possibilities for such structural
variables identified the coordinate u={\bf r}$_{OD1}\cdot$({\bf
r}$_{OD2}\times${\bf r}$_{OD3}$) as more correlated with the local
energy of the pair product wavefunction than any other combination
considered.  The distribution of u and the 
optimization deficit (see Fig.~\ref{3body}) with this pair product
wavefunction suggested the form
\begin{equation}
	\Psi_{\rm n-body}= e^{-a(u-b)^{2}}+e^{-a(u+b)^{2}}\enspace.
\end{equation}
The pair product wavefunction with moving and scaling parameters times
this new multiplicative many-body factor was optimized using Variational
Monte Carlo.  Zero point energies of 0.02838(4) Hartree
(6229(9)~/cm) and 0.03326(2) Hartree (7300(4)~/cm) are obtained for
the correlated wavefunctions with the OSS and SSH potentials,
respectively.  This is nearly as good as the VMC result with mixing of
9 states.  The optimized parameters are shown in
Table~\ref{3BodyWaveFunction}.  The new distribution and optimization
deficit for the OSS potential are shown in Fig.~\ref{3bodyOpt}.

Using pair product wavefunctions with this correlation piece as the
basis in a McMillan VMC calculation with 9 functions reduced the
ground state energy to 5860(2)~/cm and 7041(3)~/cm for OSS and SSH
potentials, respectively.  These energies contain 5-6\% error with
respect to the DMC ground state energies.

\section*{Conclusion}

A pair product wavefunction for D$_{3}$O$^{+}$ has been obtained from
an approximate density matrix.  This pair product wavefunction
captures two body interactions fairly well.  Variational Monte Carlo
includes correlation by mixing in excited states.  However, it takes
about 99 states to get to an error of 4-6\%.  Density Matrix Monte
Carlo gets accurate energies with 9 wavefunctions.  However, it does
not provide a simple wavefunction.

A compact ground state function is obtained by using the variational
principle to identify the most significant many-body terms and
including it directly into the wavefunction.  For the D$_{3}$O$^{+}$
ion the most significant such variable was the vector triple product
{\bf r}$_{OD1}\cdot$({\bf r}$_{OD2}\times${\bf r}$_{OD3}$).
Variational Monte Carlo with 9 of these correlated functions yields a
ground state wavefunction with an error of 5-6\% in the zero point
energy.
 
\section*{Acknowledgments} 

We thank the group L. Ojamae, I. Shavitt, and S. J. Singer, and the
group F. H. Stillinger, D. K. Stillinger, and J. A. Hodgdon for
pre-publication release of their potential energy surfaces.  We also
thank R. A. Harris and R. B.  Walker for helpful discussions.  This
study was supported by the LDRD program and the Center for Nonlinear Studies
at Los Alamos National
Laboratory.

\pagebreak
%
%

\pagebreak

\pagebreak

%
%

\begin{table}
\caption{Comparison of SSH and OSS potentials with B3LYP
hybrid density functional theory calculations of Martin, Hay, and
Pratt.$^{\rm a}$}
\vspace{0.1in}
\begin{tabular}{|lccc|}\hline
Quantity & SSH & OSS & {\it ab initio\/}$^{\rm a}$\cr
\hline
H+ Binding on H2O & 0.269~Hartree & 0.1695~Hartree & 0.27367~Hartree \cr
H3O+ Inversion Barrier & 4.78 kcal/mole & 4.42 kcal/mole & 2 kcal/mole \cr
zero-point (Normal Modes) & 0.0339~Hartree & 0.0281~Hartree & 0.0212~Hartree \cr
zero-point (DMC) & 0.03042~Hartree & 0.02516~Hartree &  \cr
\hline
\end{tabular}
\vspace{0.1in}
$^{\rm a}$ R. L. Martin, P. J. Hay, and L. R. Pratt. J. Phys. Chem. A 
{\bf 102}, 3565 (1998). 
\label{CompPot}
\end{table}

\begin{table}
\caption{Variational Energies for the first 5 states using 9, 17, 46, 
and 99 functions for the OSS potential.  An independent estimate for
the zero-point energy of $5522(7)$/cm is found using diffusion Monte
Carlo.  64 trajectories generating 10,000 configurations were used in 
the integrals.}
\vspace{0.1in}
	\begin{tabular}{|c|c|c|c|c|}
		\hline
        States & 9 & 17 & 46 & 99 \\
		\hline
		0 & 6198(2) & 6083(4) & 5903(4) & 5856(4) \\
		\hline
		1 & 7638(4) & 7213(6) & 7001(6) &  6905(5) \\
		\hline
		2 & 9588(5) & 9106(10) & 8702(12) & 8508(10)  \\
		\hline
		3 & 10045(6) & 9449(12) & 8804(8) & 8672(7)  \\
		\hline
		4 & 10176(6) & 9677(8) & 9269(8) & 9105(8)  \\
		\hline
	\end{tabular}
\label{OSSCFMCtau=0}
\end{table}

\begin{table}
\caption{Variational Energies for the first 5 states using 9, 17, 44, 
and 94 excited state functions for the SSH potential.  An independent
estimate for the zero-point energy of $6676(7)$/cm is found using
diffusion Monte Carlo.  64 trajectories generating 10,000 configurations 
were used in the integrals.}
\vspace{0.1in}
	\begin{tabular}{|c|c|c|c|c|}
		\hline
        States & 9 & 17 & 44 & 94 \\
		\hline
		0 & 7241(4) & 7112(4) & 6974(4) & 6939(4)\\
		\hline
		1 & 8692(6) & 8201(6) & 8065(6) & 7968(6)\\
		\hline
		2 & 10604(6) & 10176(10) & 9790(8) & 9538(8)\\
		\hline
		3 & 11498(14) & 10746(16) & 10043(8) & 9885(8)\\
		\hline
		4 & 11927(8) & 11386(8) & 10729(8) & 10484(6)\\
		\hline
	\end{tabular}
\label{SSHCFMCtau=0}
\end{table}

\begin{table}
\caption{Density matrix Monte Carlo energies for the first few states 
using 9 and 17 symmetric excited state functions for OSS. One million 
configurations where used in density matrix Monte Carlo.  The time 
step was 1~/~Hartree.  Error bars are 90\% confidence intervals.  An 
independent estimate for the zero-point energy of $5522(7)$/cm is 
found using diffusion Monte Carlo.} 
\vspace{0.1in}
	\begin{tabular}{|c|c|c|}
		\hline
        States & 9 & 17  \\
		\hline
		0 & 5525(4) & 5513(5)  \\
		\hline
		1 & 6297(10) & 6307(5) \\
		\hline
		2 & 6883(22) & 6881(28) \\
		\hline
		3 & 7979(10) &  7790(38) \\
		\hline
		4 & 8083(18) &  8038(20) \\
	\end{tabular}
\label{OSSCFMCtauplateau}
\end{table}

\begin{table}
\caption{Density matrix Monte Carlo energies for the first few states 
using 9 and 17 symmetric excited state functions for SSH. One million
configurations where used in density matrix Monte Carlo.  The time
step was 1~/~Hartree.  Error bars are 90\% confidence intervals.  Some
data was not well approximated by a decaying exponential.  In this
case, the point in the plateau with the smallest standard deviation is
given.  For this data which is designated by a *, 2 standard
deviations are given in parenthesis.  An independent estimate for the
zero-point energy of $6676(7)$/cm is found using diffusion Monte
Carlo.}
\vspace{0.1in}
\begin{tabular} [t]{|r|r|r|}
		\hline
        States & 9 & 17  \\
		\hline
		0 & 6661(5) & 6675(4) \\
		\hline
		1 & 7437(7) & 7406(7)  \\
		\hline
		2 & 7964(54)* & 7934(78)*   \\
		\hline
		3 & 8912(4) & 8918(10)   \\
		\hline
		4 & 10119(25) & 9061(43)*  \\
		\hline
	\end{tabular}
\label{SSHCFMCtauplateau}
\end{table}

\begin{table}
\caption{Parameters used for the variational ground state wavefunction.}
\vspace{0.1in}
	\begin{tabular}{|c|c|c|}
		\hline
         & OSS & SSH  \\
		\hline
		s$_{OD}$ & 0.911 & 0.853  \\
		\hline
		s$_{DD}$ & 0.935 & 0.895 \\
		\hline
		m$_{OD}$ & 0.157 & 0.242 \\
		\hline
		m$_{DD}$ & 0.078 & 0.100 \\
		\hline
		a & 9. & 10. \\
		\hline
		b & 0.851 &  0.911 \\
	\end{tabular}
\label{3BodyWaveFunction}
\end{table}


\begin{thebibliography}{10}

\bibitem{Stillinger:78}
F.~H. Stillinger and C.~W. David, J. Chem. Phys. {\bf 69},  1475  (1978).

\bibitem{Morse:81}
M.~D. Morse and S.~A. Rice, J. Chem. Phys. {\bf 74},  6514  (1981).

\bibitem{Christou:81}
N.~I. Christou, J.~S. Whitehouse, D. Nicholson, and N.~G. Parsonage, Faraday
  Sym. Chem. Soc.  139  (1981).

\bibitem{Stillinger:82}
T.~A. Weber and F.~H. Stillinger, J. Phys. Chem. {\bf 86},  1314  (1982).

\bibitem{Reimers:82}
J.~R. Reimers, R.~O. Watts, and M.~L. Klein, Chem. Phys. {\bf 64},  95  (1982).

\bibitem{Berens}
P.~H. Berens, D.~H.~J. Mackay, G.~M. White, and K.~R. Wilson, J. Chem. Phys.
  {\bf 79},  2375  (1983).

\bibitem{Bopp:83}
P. Bopp, G. Jancso, and K. Heinzinger, Chem. Phys. Letts. {\bf 98},  129
  (1983).

\bibitem{Jancso}
G. Jancso and P. Bopp, Z. Naturforsch. {A} {\bf 38},  206  (1983).

\bibitem{Teleman:83}
O. Teleman, B. Jonsson, and S. Engstrom, Molec. Phys. {\bf 60},  193  (1987).

\bibitem{Thuraisingham:83}
R.~A. Thuraisingham and H.~L. Friedman, J. Chem. Phys. {\bf 78},  5772  (1983).

\bibitem{Deutsch}
P.~W. Deutsch and T.~D. Stanik, J. Chem. Phys. {\bf 85},  4660  (1986).

\bibitem{Wojcik}
M.~J. Wojcik, J. Molec. Struct. {\bf 189},  89  (1988).

\bibitem{Wallqvist:90}
A. Wallqvist, Chem. Phys. {\bf 148},  439  (1990).

\bibitem{Ruff}
I. Ruff and D.~J. Diestler, J. Chem. Phys. {\bf 93},  2032  (1990).

\bibitem{Slanina:90}
Z. Slanina, Chem. Phys. Letts. {\bf 172},  367  (1990).

\bibitem{Wallqvist:91}
A. Wallqvist and O. Teleman, Molec. Phys. {\bf 74},  515  (1991).

\bibitem{Zhu}
S.~B. Zhu, S. Yao, J.~B. Zhu, S. Singh, and G.~W. Robinson, J. Phys. Chem. {\bf
  95},  6211  (1991).

\bibitem{Suhm:91}
M.~A. Suhm and R.~O. Watts, Molec. Phys. {\bf 73},  463  (1991).

\bibitem{Ojamae:92b}
L. Ojamae, K. Hermansson, and M. Probst, Chem. Phys. Letts. {\bf 191},  500
  (1992).

\bibitem{Ojamae:92a}
L. Ojamae, J. Tegenfeldt, J. Lindgren, and K. Hermansson, Chem. Phys. Letts.
  {\bf 195},  97  (1992).

\bibitem{Corongiu:92}
G. Corongiu, Int. J. Quant. Chem. {\bf 42},  1209  (1992).

\bibitem{Smith:92}
D.~E. Smith and A.~D.~J. Haymet, J. Chem. Phys. {\bf 96},  8450  (1992).

\bibitem{Halley}
J.~W. Halley, J.~R. Rustad, and A. Rahman, J. Chem. Phys. {\bf 98},  4110
  (1993).

\bibitem{Corongiu:93a}
G. Corongiu and E. Clementi, J. Chem. Phys. {\bf 98},  4984  (1993).

\bibitem{Sciortino}
F. Sciortino and G. Corongiu, J. Chem. Phys. {\bf 98},  5694  (1993).

\bibitem{Trokhymchuk:93}
A.~D. Trokhymchuk, M.~F. Holovko, and K. Heinzinger, J. Chem. Phys. {\bf 99},
  2964  (1993).

\bibitem{Vossen:94}
M. Vossen and F. Forstmann, J. Chem. Phys. {\bf 101},  2379  (1994).

\bibitem{Mizan}
T.~I. Mizan, P.~E. Savage, and R.~M. Ziff, J. Phys. Chem. {\bf 98},  13067
  (1994).

\bibitem{Duh}
D.~M. Duh, D.~N. Perera, and A.~D.~J. Haymet, J. Chem. Phys. {\bf 102},  3736
  (1995).

\bibitem{Kalinichev}
A.~G. Kalinichev and K. Heinzinger, Geochim. Cosmochim. Acta {\bf 59},  641
  (1995).

\bibitem{David:96}
C.~W. David, J. Chem. Phys. {\bf 104},  7255  (1996).

\bibitem{Bansil}
R. Bansil, T. Berger, K. Toukan, M.~A. Ricci, and S.~H. Chen, Chem. Phys.
  Letts. {\bf 132},  165  (1986).

\bibitem{Silvestrelli}
P.~L. Silvestrelli, M. Bernasconi, and M. Parrinello, Chem. Phys. Letts. {\bf
  277},  478  (1997).

\bibitem{Rick}
Y. Liu, K. Kim, B.~J. Berne, R.~A. Friesner, and S.~W. Rick, to be published  .

\bibitem{OSS}
L. Ojamae, I. Shavitt, and S.~J. Singer, to be published  .

\bibitem{Tuckerman:97}
M.~E. Tuckerman, D. Marx, M.~L. Klein, and M.~. Parrinello, Science {\bf 275},
  817  (1997).

\bibitem{Pomes:95}
R. Pom\'es and B. Roux, Chem. Phys. Lett. {\bf 234},  416  (1995).

\bibitem{tuckerman:95}
M. Tuckerman, K. Laasonen, M. Sprik, and M. Parrinello, J. Phys. Chem. {\bf
  99},  5749  (1995).

\bibitem{Pomes:96}
R. Pom\'es and B. Roux, J. Phys. Chem {\bf 100},  2519  (1996).

\bibitem{HLR}
B.~L. Hammond, W.~A. Lester, Jr., and P.~J. Reynolds, {\em Monte Carlo Methods
  in Ab Initio Quantum Chemistry} (World Scientific, River Edge, NJ, USA,
  1994).

\bibitem{wlm}
W.~L. McMillan, Phys. Rev. A {\bf 138},  442  (1965).

\bibitem{ceperley:88}
D.~M. Ceperley and B. Bernu, J. Chem. Phys. {\bf 89},  6316  (1988).

\bibitem{bernu:90}
B. Bernu, D.~M. Ceperley, and W.~A. Lester, Jr., J. Chem. Phys. {\bf 93},  552
  (1990).

\bibitem{Brown:95}
W.~R. Brown, W.~A. Glauser, and W.~A. Lester, Jr., J. Chem. Phys. {\bf 103},
  9721  (1995).

\bibitem{feynman}
R.~P. Feynman, {\em Statistical Mechanics, A Set of Lectures} (Benjamin,
  Reading, Massachusetts, 1972), see chapter 3.

\bibitem{Singh}
A. Sethia, S. Sanyal, and Y. Singh, J. Chem. Phys. {\bf 93},  7268  (1990).

\bibitem{metropolis}
N. Metropolis, A.~W. Rosenbluth, M.~N. Rosenbluth, A.~H. Teller, and E. Teller,
  J. Chem. Phys. {\bf 21},  1087  (1953).

\bibitem{kalos}
M.~H. Kalos and P.~A. Whitlock, {\em Monte Carlo Methods, Volume I: Basics}
  (John Wiley and Sons, New York, 1986).

\bibitem{SSHpot}
F.~H. Stillinger, D.~K. Stillinger, and J.~A. Hodgdon, preprint  .

\bibitem{NumericalRecipies}
W.~H. Press, S.~A. Teukolsky, W.~T. Vetterling, and B.~P. Flannery, {\em
  Numerical Recipies in C: The Art of Scientific Computing, Second Edition}
  (Cambridge University Press, Cambridge, 1992), pp.\ 113--116.

\bibitem{Sears}
T.~J. Sears, P.~R. Bunker, P.~B. Davies, S.~A. Johnson, and V. \v{S}pirko, J.
  Chem. Phys. {\bf 83},  2676  (1985).

\end{thebibliography}
\end{document}